\documentclass[twocolumn]{aastex631}
\usepackage{stfloats}
\usepackage{float}
\usepackage{graphicx}
\usepackage{natbib}
\usepackage{hyperref}
\usepackage{amsmath}
\usepackage{autobreak}
\usepackage{multirow}
\usepackage{makecell}
\usepackage{color}
\usepackage{bm}
\usepackage{threeparttable}
\usepackage[figuresright]{rotating}
\usepackage[mathscr]{euscript}
\usepackage[utf8]{inputenc}
\usepackage[T1]{fontenc}
\usepackage{amsmath}
\usepackage{makecell}

\graphicspath{{./}{figures/}}

\begin{document}

\title{The dusty red supergiant progenitor and the local environment of the Type~II SN~2023ixf in M101}

\correspondingauthor{Ning-Chen Sun}
\email{sunnc@ucas.ac.cn}

\author[0000-0002-3651-0681]{Zexi Niu}
\affiliation{School of Astronomy and Space Science, University of Chinese Academy of Sciences, Beijing 100049, People's Republic of China}
\affiliation{National Astronomical Observatories, Chinese Academy of Sciences, 20A Datun Road, Chaoyang District, Beijing, China}

\author[0000-0002-4731-9698]{Ning-Chen Sun}
\affiliation{School of Astronomy and Space Science, University of Chinese Academy of Sciences, Beijing 100049, People's Republic of China}
\affiliation{National Astronomical Observatories, Chinese Academy of Sciences, 20A Datun Road, Chaoyang District, Beijing, China}

\author[0000-0003-0733-7215]{Justyn R. Maund}
\affiliation{Department of Physics and Astronomy, University of Sheffield, Hicks Building, Hounsfield Road, Sheffield, S3 7RH, United Kingdom}

\author[0000-0003-2536-2641]{Yu Zhang}
\affiliation{Key Laboratory of Optical Astronomy, National Astronomical Observatories, Chinese Academy of Sciences, Beijing 100101, China}

\author[0000-0003-4936-4959]{Ruining Zhao}
\affiliation{School of Astronomy and Space Science, University of Chinese Academy of Sciences, Beijing 100049, People's Republic of China}
\affiliation{National Astronomical Observatories, Chinese Academy of Sciences, 20A Datun Road, Chaoyang District, Beijing, China}

\author{Jifeng Liu}
\affiliation{New Cornerstone Science Laboratory, National Astronomical Observatories, Chinese Academy of Sciences, Beijing 100012, People's Republic of China}
\affiliation{School of Astronomy and Space Science, University of Chinese Academy of Sciences, Beijing 100049, People's Republic of China}
\affiliation{Institute for Frontiers in Astronomy and Astrophysics, Beijing Normal University, Beijing, 102206, People's Republic of China}

\begin{abstract}

As one of the closest supernovae (SNe) in the last decade, SN~2023ixf is an unprecedented target to investigate the progenitor star that exploded.
However, there is still significant uncertainty in the reported progenitor properties.
In this work, we present a detailed study of SN~2023ixf's progenitor with two independent analyses.
We first modelled its spectral energy distribution (SED) based on Hubble Space Telescope optical, Spitzer mid-infrared (IR), and ground-based near-IR data.
We find that stellar pulsation and circumstellar extinction have great impacts on SED fitting, and the result suggests a relatively massive red supergiant (RSG) surrounded by C-rich dust with an initial mass of 16.2--17.4 $M_{\odot}$.
The corresponding rate of mass-loss occurring at least 3 years before the SN explosion is about $2 \times 10^{-4} M_\odot$yr$^{-1}$.
We also derived the star formation history of the SN environment based on resolved stellar populations, and the most recent star-forming epoch corresponds to a progenitor initial mass of 17--19~$M_\odot$, in agreement with that from our SED fitting. 
Therefore, we conclude that the progenitor of SN~2023ixf is close to the high-mass end for Type II SN progenitors.

\end{abstract}

\section{Introduction}

Core-collapse supernovae (SNe) are the spectacular explosions of dying massive ($>$8~$M_\odot$) stars. It is a major goal, and currently a major difficulty, to determine the progenitor stars of different SN types. It has been confirmed that the Type~II-P SNe arise from the explosion of red supergiants (RSGs) with almost 20 directly probed progenitors (e.g. SN~2003gd, \citealt{Smartt2004, Maund2009}; SN~2005cs, \citealt{Maund2005}; SN~2017eaw, \citealt{2019VanDyk, Rui2019}). However, none of them appear more massive than $\sim$16--18~$M_\odot$, which is significantly lower than the theoretically predicted upper mass limit of 25--30~$M_\odot$ (i.e. the ``RSG problem"; \citealt{2009Smartt, 2015Smartt}; although see also \citealt{2018Davies}). This could be due to the very uncertain circumstellar extinction, which may lead to underestimation of the progenitor masses \citep{Walmswell2012}; it is also possible that the more massive stars may explode as other types of SNe (e.g. Type~II-L or Type~IIn; \citealt{Groh2013}) or even collapse directly into a black hole \citep{Sukhbold2016}.

SN~2023ixf is a Type~II SN that recently exploded in the nearby galaxy M101 \citep[i.e. the Whirpool galaxy;][]{Itagaki2023}.  
It serves an unprecedented example of studying properties of the progenitor. 
Soon after the explosion, the stellar variability of the progenitor in Infrared (IR) band was identified by \citet{2023Szalai}
Simultaneously, several groups reported the detection of a progenitor candidate for SN~2023ixf on pre-explosion images \citep[e.g.][]{Pledgerixf, Kilpatrickixf, Jencsonixf, Soraisamixf}. 
Their results are all consistent with a RSG progenitor enshrouded by a dusty envelope.  
The presence of dense circumstellar material (CSM) around the progenitor is also indicated by the fast rising luminosity of the SN and its early spectra with prominent and narrow nebular emission lines \citep{Yamanaka2023, 2023Smith, Vasylyev2023, Teja2023, 2023JacobsonGalan, Hiramatsu2023, 2023Bostroem,2023Hosseinzadeh}.
However, the inferred progenitor mass is still under debate and can range from 8--10~$M_\odot$ up to $\sim$20~$M_\odot$. 
This large uncertainty presents a significant challenge to further studies of SN~2023ixf.

In this paper, we carry out a detailed analysis of the progenitor of SN~2023ixf in order to derive its accurate properties. 
We use two different techniques, based on the direct progenitor detection and SN environment, to assess the reliability of our results against possible systematic errors. Throughout this paper, we adopt a distance of 6.85 $\pm$~0.15~Mpc \citep{2022Riess}, a Milky Way extinction of $E(B - V)_{\rm MW} = 0.008$ mag \citep{2011sfd}, a host galaxy extinction of $E(B - V)_{\rm host} = 0.033$ \citep{2023Lundquist,2023Smith,2023JacobsonGalan}, and a standard extinction law with $R_V = 3.1$ \citep{ebvlaw.ref}. All magnitudes are reported in the Vega system unless otherwise specified.

\section{Data and photometry}\label{sec.phot}

\begin{figure*} 
    \centering 
    \includegraphics[width=0.8\linewidth]{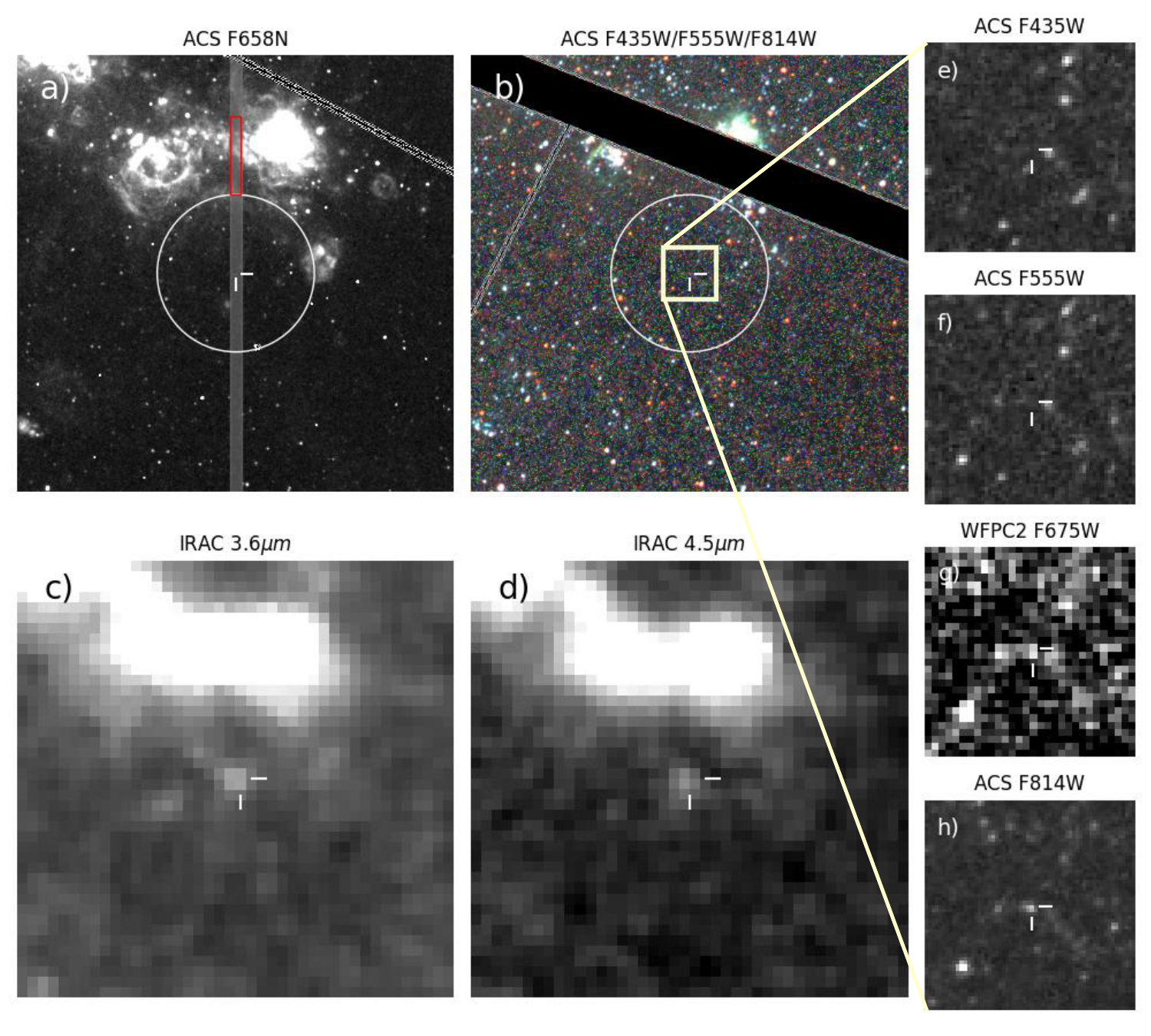}
    \caption{(a, b) HST and (c, d) Spitzer images of the site of SN~2023ixf. The SN position is shown by the cross-hair and the images have a dimension of 25~arcsec $\times$ 25~arcsec. In the F658N narrow-band image (a), a few H~\textsc{ii} regions are visible to the west and to the north of SN~2023ixf, and the grey-shaded stripe shows the slit position in one of our long-slit spectroscopic observations of the SN (Zhang et al. in preparation); we extracted a spectrum from the red-outlined area in order to estimate the metallicity of the ionized gas (Section~\ref{sec.chemical}). In (a) and (b), the circle is centered at the SN position with a radius of 150~pc, within which we selected stars for an environmental analysis (Section~\ref{sec.pop}). (e-h) Stamps of HST images in the F435W, F555W, F675W, and F814W bands, all with a dimension of 3~arcsec $\times$ 3~arcsec. All panels are aligned North up and East to the left.
  \label{prephot}}
\end{figure*}

\begin{figure} 
    \hspace{-0.5cm}
    \includegraphics[width=1.1\linewidth]{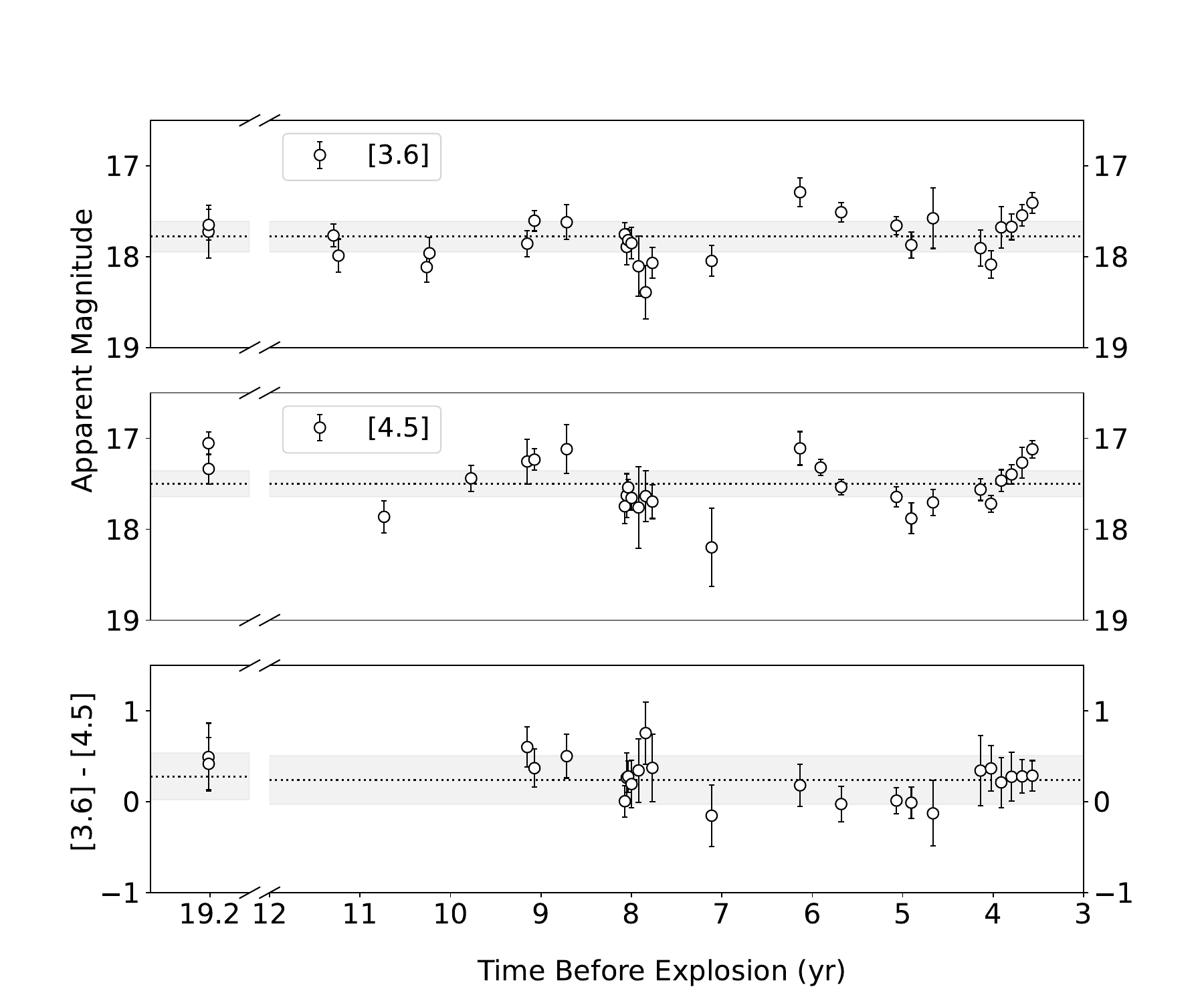}
    \caption{Light and color curves of the progenitor of SN~2023ixf. The dotted horizontal lines are the phase-weighted averages and the light-shaded regions reflect the typical photometric uncertainties. \label{fig.IRACphot}} 
\end{figure}

\subsection{HST optical data}

The site of SN~2023ixf was observed by Hubble Space Telescope (HST) with the Wide Field and Planetary Camera 2 (WFPC2), the Wide Field Camera 3 (WFC3), and the Advanced Camera for Surveys (ACS) before explosion. We retrieved the calibrated images from the Mikulski Archive for Space Telescopes\footnote{\url{https://archive.stsci.edu}} and reprocessed them with the \textsc{astrodrizzle} package \citep{2012Gonzaga} for better image alignment and cosmic ray removal. On the ACS F814W image, there are two objects within 0.2~arcsec from the reported SN position and \citet{Kilpatrickixf} identified the relatively brighter one as the SN progenitor (Fig.~\ref{prephot}) based on differential astrometry with post-explosion images. We performed point-spread-function (PSF) photometry with the \textsc{dolphot} package \citep{2000Dolphin} and detected the progenitor significantly in the F658N, F675W, and F814W bands. The measured magnitudes, which are listed in Table~\ref{HSTphot}, are slightly brighter than those reported by \citet{Kilpatrickixf} within 1--4$\sigma$ uncertainties (note their magnitudes are in the AB system). The difference could be due to our different parameters in \textsc{dolphot} photometry, and this difference does not have any significant effect in the following analysis. The progenitor was not detected in any other bands, and we estimated the detection limits with artificial star tests. We also used the ACS F435W, F555W, and F814W photometry to perform an environmental analysis of the resolved stellar populations around SN~2023ixf (Section~\ref{sec.pop}).

\begin{table} [H]
    \centering
    \caption{HST observations used in this work and photometry of SN~2023ixf's progenitor. The other HST observations are not listed since their detection limits are not very constraining for the progenitor's SED (Section~\ref{sec.sedfitting}) or because they are not used in the environmental analysis (Section~\ref{sec.pop}).\label{HSTphot}}
    \begin{tabular}{cccc}
    \hline
    \hline
    Epoch & Program & Instrument & Magnitude \\
    (MJD) & ID & and filter & \\
\hline
    52593.99 & 9490$^{\rm a}$ & ACS/F435W  &  $>27.3$ \\
    52594.01 & 9490 & ACS/F555W  &  $>27.0$ \\
    53045.01 & 9720$^{\rm b}$ & ACS/F658N  &   24.75 (0.20) \\
    51261.04 & 6829$^{\rm c}$ & WFPC2/F675W  &   25.36 (0.19) \\
    52594.02 & 9490 & ACS/F814W  &  24.34 (0.05) \\
\hline
    \end{tabular} \\
    PIs: (a) K. Kuntz; (b) P. Chandar; (c) Y.-H. Chu.
\end{table}

\subsection{Spitzer mid-IR data}\label{sec.spitzer}

The site of SN~2023ixf was observed by Spitzer/IRAC at a total of 31 epochs during its cold mission in 2004 and during the warm mission in 2012--2019. We retrieved the post-basic calibrated (PBCD) images from the Spitzer Heritage Archive\footnote{\url{https://sha.ipac.caltech.edu}}, and a pre-explosion source is clearly visible at the SN position in the [3.5] and [4.5] bands (Fig.~\ref{prephot}). We used the \textsc{dophot} package \citep{dophot} to perform PSF photometry at individual epochs, the results of which are listed in Table~\ref{IRACphot} and displayed in Fig.~\ref{fig.IRACphot}. Prominent variability, with a semi-amplitude of $\sim$0.25 mag, can be seen in both the [3.6] and [4.5] bands without obvious color variation. This variability is significant compared with the photometric uncertainties and not due to the possible zero-point shift between different epochs (using non-variable field stars to provide a reference, we found the zero-point shift, if any, to be negligible compared with the progenitor's variability). This variability was also reported in \citet{Kilpatrickixf}, \citet{Jencsonixf}, and \citet{Soraisamixf}, and they all found a pulsational period of $\sim$1000 days. We derived phase-weighted average magnitudes of [3.6] = 17.78 $\pm$~0.19~mag and [4.5] = 17.50 $\pm$~0.20~mag. On the [5.8] and [8.0] images, however, no counterpart can be significantly detected at the SN position down to 3$\sigma$ detection limits of 14.95 and 14.38~mag, respectively.


\begin{table} [htbp]
    \centering
    \caption{Spitzer/IRAC photometry of the progenitor of SN~2023ixf in the [3.6] and [4.5] bands.\label{IRACphot}}
    \begin{tabular}{ccc}
    \hline
    \hline
Epoch & [3.6] & [4.5] \\
(MJD) & (mag) & (mag) \\
\hline
 53072.09 & 17.73 (0.29) & 17.05 (0.12) \\ 
 53072.49 & 17.65 (0.17) & 17.34 (0.17) \\ 
 55960.72 & 17.77 (0.13) & --  \\ 
 55980.99 & 17.99 (0.18) & --  \\ 
 56165.01 & --  & 17.86 (0.17) \\ 
 56337.07 & 18.12 (0.17) & --  \\ 
 56348.11 & 17.96 (0.17) & --  \\ 
 56516.35  & --  & 17.44 (0.15) \\ 
 56742.84 & 17.86 (0.15) & 17.25 (0.25) \\ 
 56771.83 & 17.60 (0.11) & 17.23 (0.12) \\ 
 56902.01 & 17.62 (0.19) & 17.12 (0.27) \\ 
 57136.69 & 17.75 (0.13) & 17.75 (0.19) \\ 
 57144.06 & 17.89 (0.20) & 17.63 (0.24) \\ 
 57150.17 & 17.82 (0.11) & 17.54 (0.09) \\ 
 57163.71 & 17.85 (0.17) & 17.66 (0.13) \\ 
 57191.82 & 18.11 (0.33) & 17.76 (0.45) \\ 
 57220.79 & 18.39 (0.29) & 17.64 (0.28) \\ 
 57247.82 & 18.07 (0.17) & 17.70 (0.19) \\ 
 57486.85 & 18.05 (0.17) & 18.20 (0.43) \\ 
 57843.93 & 17.29 (0.16) & 17.11 (0.18) \\ 
 57926.90 & -- & 17.32 (0.09) \\ 
 58009.67 & 17.51 (0.11) & 17.54 (0.09) \\ 
 58232.95 & 17.66 (0.10) & 17.64 (0.11) \\ 
 58292.87 & 17.87 (0.14) & 17.88 (0.17) \\ 
 58380.22 & 17.58 (0.33) & 17.70 (0.14) \\ 
 58572.08 & 17.91 (0.20) & 17.56 (0.12) \\ 
 58614.39 & 18.09 (0.15) & 17.72 (0.09) \\ 
 58655.68 & 17.68 (0.23) & 17.46 (0.12) \\ 
 58697.50 & 17.67 (0.14) & 17.40 (0.11) \\ 
 58740.01 & 17.55 (0.12) & 17.27 (0.17) \\ 
 58781.31 & 17.41 (0.12) & 17.12 (0.09) \\ 
 \hline
phase-weighted   \\
average & 17.78 (0.19) & 17.50 (0.20) \\
 \hline
\end{tabular}
\end{table}

\subsection{Ground-based near-IR data}

During the writing of this paper, \citet{Soraisamixf} reported $JHK$-band light curves acquired from the Gemini Near-IR Imager (NIRI) and the Wide Field Camera (WFCAM) on the United Kingdom Infrared Telescope (UKIRT) images.
Their photometry is roughly consistent with those derived by \citet{Jencsonixf} and \citet{Kilpatrickixf} based on observations with NIRI, NEWFIRM infrared camera, and/or the MMT and Magellan Infrared Spectrograph (MMIRS). 
In this paper, we applied $JHK$-band photometry of \citet{Soraisamixf} and computed the phase-weighted average $J$ = 20.67 $\pm$~0.19~mag,  $H$ = 19.55 $\pm$ 0.13~mag, and $K$ = 18.69 $\pm$~0.08~mag with their period and amplitudes, since their data have better phase coverage.

\section{SED analysis}\label{sec.sedfitting}

\begin{figure*} [htbp] 
    \centering
    \includegraphics[width=1.0\linewidth]{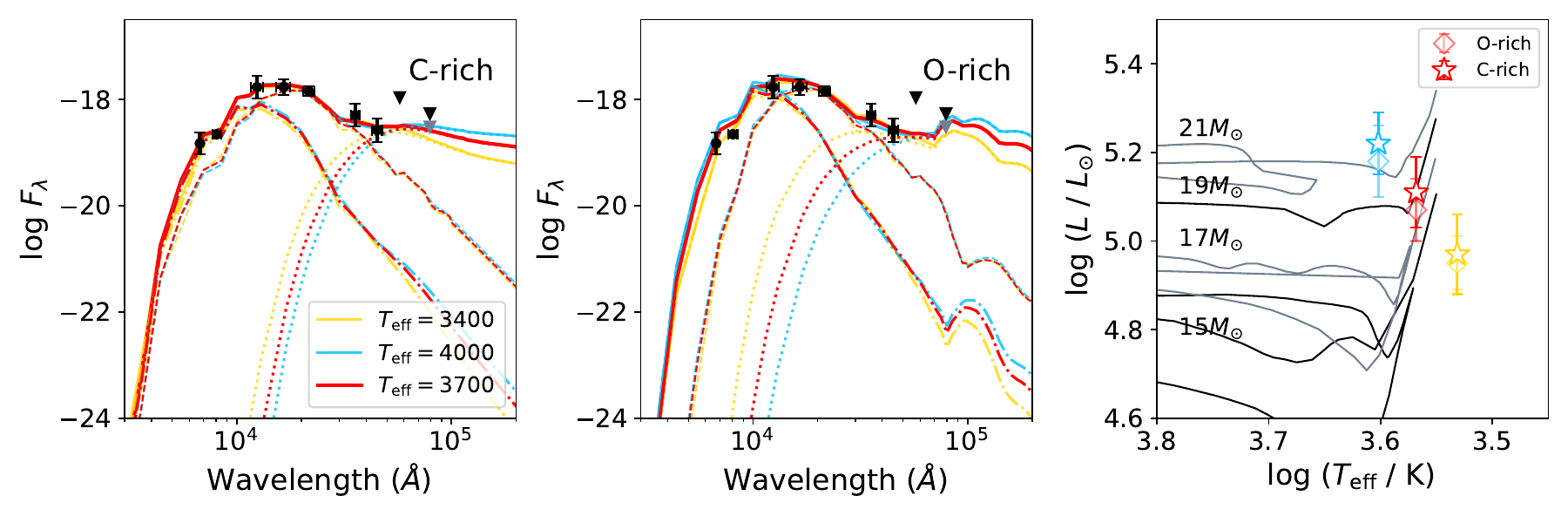}
    \caption{
(a) Observed SED (black data points) and the best-fitting model with C-rich dust (colored solid lines) for the progenitor of SN~2023ixf. The vertical error bars reflect their 3$\sigma$ photometric uncertainties and the horizontal error bars are the root-mean-square bandwidths. Only 3$\sigma$ detection limits in Spitzer/IRAC [5.8] and [8.0] bands are displayed (inverted triangles), and those in the other bands are not very constraining on the progenitor's SED. The 2$\sigma$ detection limit in [8.0] band is shown in gray. 
The dashed, dotted, and dot-dashed lines correspond to the attenuated RSG radiation, dust emission, and dust-scattered radiation, respectively. 
The yellow, red, and blue colors correspond to effective temperatures of $T_{\rm eff}$ = 3400, 3700, and 4000~K for the RSG progenitor. Detailed model parameters are listed in Table \ref{tab.para}. (b) Same as (a) but for the O-rich dust model. Notice that the prominent silicate bumps of $T_{\rm eff}$ = 3700 and 4000~K models almost exceed the [8.0] detection limit. 
(c) The progenitor of SN~2023ixf on the Hertzsprung-Russell diagram. The  diamonds and stars demonstrate models with O-rich and C-rich dust, respectively. They are colored in the same way as (a-b). 
Overlaid black/grey lines are the \textsc{parsec} stellar evolutionary tracks for different initial masses.
    \label{hrd}} 
\end{figure*}

\begin{table*} [htbp]
    \centering
    \caption{Best-fitting and derived parameters of the SED for the progenitor of SN~2023ixf .\label{tab.para}}
    \begin{tabular}{ccccccc}
    \hline
    \hline
       & $T_{\rm eff}$ & log($L/L_{\odot}$) & $T_{\rm in}$ & $\tau_V$ & $E(B-V)_{\rm CSM}$ & Comment \\ 
       & (K) & & (K) & & (mag) & \\
       \hline
    C-rich
    & 3400 & 4.97$_{-0.08}^{+0.09}$ & 514$_{-70}^{+96}$ & 5.37$_{-0.59}^{+0.64}$ & 1.39 & \\
    & 3700 & 5.11$_{-0.08}^{+0.08}$ & 433$_{-50}^{+55}$ & 6.39$_{-0.59}^{+0.58}$ & 1.64 & (a)\\
    & 4000 & 5.22$_{-0.08}^{+0.07}$ & 387$_{-39}^{+46}$ & 7.16$_{-0.53}^{+0.54}$ & 1.85 & \\
    \hline
    O-rich
    & 3400 & 4.95$_{-0.05}^{+0.06}$ & 1025$_{-283}^{+317}$ & 12.13$_{-0.93}^{+0.86}$& 1.88 & \\
    & 3700 & 5.07$_{-0.06}^{+0.08}$ & 715$_{-208}^{+337}$ & 12.79$_{-0.93}^{+0.94}$ & 1.98 & (b) \\
    & 4000 & 5.18$_{-0.05}^{+0.09}$ & 553$_{-298}^{+272}$ & 13.15$_{-0.91}^{+1.01}$ & 2.04 & (b) \\
    \hline
    \end{tabular} \\
    Comments: (a) This is the most favored effective temperature since the other temperatures are either too cold or too warm compared with that of a RSG just before the explosion (Fig.~\ref{hrd}c). (b) These two models have prominent silicate bumps that almost exceed the [8.0] detection limits. 
\end{table*}

\subsection{Method}

RSGs could experience significant mass loss and enshroud themselves within dusty envelopes \citep{1988deJager,2005van,2005Massey,2020Beasor}. The uncertain circumstellar extinction often leads to an underestimation of the progenitor mass based on detections in only one or a few filters \citep[e.g.][]{2019VanDyk}. For the progenitor of SN~2023ixf, however, the extensive optical, near-IR and mid-IR data allow us to perform a detailed modeling of its spectral energy distribution (SED).

The intrinsic SED of the SN progenitor was synthesized by the \textsc{marcs} model atmospheres \citep{2008marcs}, assuming a microtubulent velocity of 5~km~s$^{-1}$, surface gravity log($g$) = 0~dex and metallicity [Fe/H] = $-$0.25~dex (close to our estimation in Section~\ref{sec.chemical}). 
For the effective temperature, we tried three different values of $T_{\rm eff}$ = 3400, 3700, and 4000~K, which are typical of RSGs when they explode (e.g. \citealp{2015Smartt}, although see \citealp{2013Davies}). The radiation transfer through the dusty envelope was then solved with the \textsc{dusty} code. We used a $\rho \propto r^{-2}$ wind-like radial density profile and the default relative shell thickness that is 1000 times the inner radius. The standard Mathis, Rumpl \& Nordsieck (1977; MRN) grain size distribution \citep{MRN.ref} was adopted, and we considered two different types of dust grains made of either pure graphite or pure silicate \citep{1984DL}, representing a C-rich or O-rich chemical composition, respectively. The progenitor's bolometric luminosity, $L_{\rm bol}$, the dust temperature at the inner shell boundary, $T_{\rm in}$ (which is required to be lower than 1500~K such that the dust can survive in the envelope; \citealt{2004Pozzo,2018Sarangi}), and the V-band optical depth of the envelope, $\tau_V$, were left as free parameters to be fitted from the data.

We used the Markov Chain Monte Carlo method to search for models that match the progenitor detections in the F675W, F814W, J, H, K, [3.6], and [4.5] filters. We used the phase-weighted average magnitudes for the near- and mid-IR filters; for the optical F675W and F814W filters, however, data at only one epoch are available and it is difficult to estimate their phase-weighted average magnitudes. Therefore, in these two filters, we conservatively allowed the model magnitudes to vary within 0.5~mag from the observed ones, accounting for their possible variability due to stellar pulsation (see Section~\ref{sec.comparison} for a more detailed discussion).

\subsection{Results}

The best-fitting model SED with C-rich/O-rich dust and with different progenitor effective temperatures are displayed in Fig.~\ref{hrd} (a, b), and the detailed parameters are listed in Table~\ref{tab.para}. 
The circumstellar extinction $E(B-V)_{\rm CSM}$ is also computed according to the equation in \citet{2012Kochanek} and convolved with the passbands.
Note, however, the O-rich models with $T_{\rm eff} = $ 3700 and 4000 K have prominent silicate bumps that almost exceed the 3$\sigma$ [8.0] detection limit, and significantly exceed the limit at the 2$\sigma$ level (15.0~mag). Therefore, there is a 95\% probability that these two models are against the observations.

The position of SN~2023ixf's progenitor on the Hertzsprung-Russell diagram is shown in Fig.~\ref{hrd}c in comparison with the \textsc{parsec} v1.2S single-star evolutionary tracks \citep{parsec.ref}. 
Although we tried 3 different effective temperatures typical of RSGs, only the intermediate value (i.e. $T_{\rm eff}$ = 3700~K) is consistent with the end points of the tracks, while the other two values (3400 and 4000~K) are either too cold or too warm for a RSG just before the explosion (we note, however, there could be uncertainties in the estimate and model prediction of the effective temperature of RSGs; e.g. \citealp{2013Davies}.) 
Therefore, assuming single-star evolution SN~2023ixf is most likely to have a relatively massive progenitor with $M_{\rm ini}$ = 16.2--17.4~$M_\odot$ enshrouded by a C-rich dusty envelope. 

\subsection{Mass-loss rate}

Given the CSM parameters derived from SED fitting, the mass-loss rate ($\dot M$) can be calculated with 
$\dot M = \frac{16}{3}  \pi  R_{\rm in} \tau_V \rho_d  a V_{\rm exp}  r_{\rm gd} Q_{\lambda}^{-1} $ \citep{2016Beasor}. For the expansion velocity, we use $V_{\rm exp} = 115$~kms$^{-1}$ as measured by \citet{2023Smith} based on high-resolution SN spectra. For the dust grains, we assume an effective extinction efficiency of $Q_V$ = 0.4 and a bulk density of $r_{\rm gd}$ = 2.26~g~cm$^{-3}$ \citep[typical of graphites;][]{1984DL} and a grain size of $a = \sqrt{0.005\times0.25}\ \mu$m (similar to that adopted in \citealp{2020Humphreys}). Considering the half-solar metallicity of the SN environment (see Section \ref{sec.chemical}), we use a gas-to-dust ratio of 400 \citep{2005vanLoon}. With these values, we infer a mass-loss rate of $2 \times 10^{-4} M_{\odot}$yr$^{-1}$ for the CSM of the SN progenitor.

On another hand, \citeauthor{2020Beasor} (\citeyear{2020Beasor}, see also their Erratum in \citeyear{2023Beasor}) established an empirical mass-dependent $\dot M$–$L_{\rm bol}$ relation based on Galactic and LMC RSGs. Our derived mass-loss rate is significantly larger than that expected from this relation by almost 2 orders of magnitude. This suggests that the mass-loss rate is significantly enhanced for a RSG shortly before its explosion.

Meanwhile, mass-loss rates inferred from both flash spectroscopy and early light curve of SN~2023ixf are $>10^{-3} M_{\odot}$yr$^{-1}$ \citep{2023JacobsonGalan, Hiramatsu2023, 2023Bostroem}, significantly higher than that from SED modelling. 
This tension is also noticed in the previous works of \citealp{Jencsonixf}. 
We note that the dense CSM probed by the SN flash spectroscopy and early light curve was distributed within (3--7)$\times 10^{14}$ cm, while the CSM probed by the direct progenitor detection in this paper has a much larger inner radius of 1.5$\times 10^{15}$ cm. 
Also note that the pre-explosion photometry we used were taken at 3--19 years before the SN explosion. 
Therefore, it is possible that the progenitor experienced a relatively enhanced mass-loss ($\dot M \approx 2 \times 10^{-4} M_{\odot}$yr$^{-1}$) until several years before the explosion and experienced a extreme mass-loss ($\dot M = 10^{-3}-10^0 M_{\odot}$yr$^{-1}$) later during 1--3 years before the explosion.
It is also worth mentioning that the discrepancy of mass-loss rate maybe originates from nonhomogeneous CSM \citep{2023Smith,2023Beasor,2023Soker}.

In addition to continuous mass-loss, for mass-loss through eruptions, precursor outburst like that observed in normal Type II SN~2020tlf are excluded by \citet{2023Dong}.  
\citet{2023Neustadt} found that transient with peak luminosity $> 2 \times 10^{39}$~ergs$^{-1}$ cannot happen during 1--15 year before the SN explosion. Higher non-detection limits can be also seen in \citet{2023Flinner} in ultraviolet bands.

\section{Environmental analysis}\label{sec.env}

\begin{figure*}[htbp] 
    \centering 
    \includegraphics[width=0.9\linewidth]{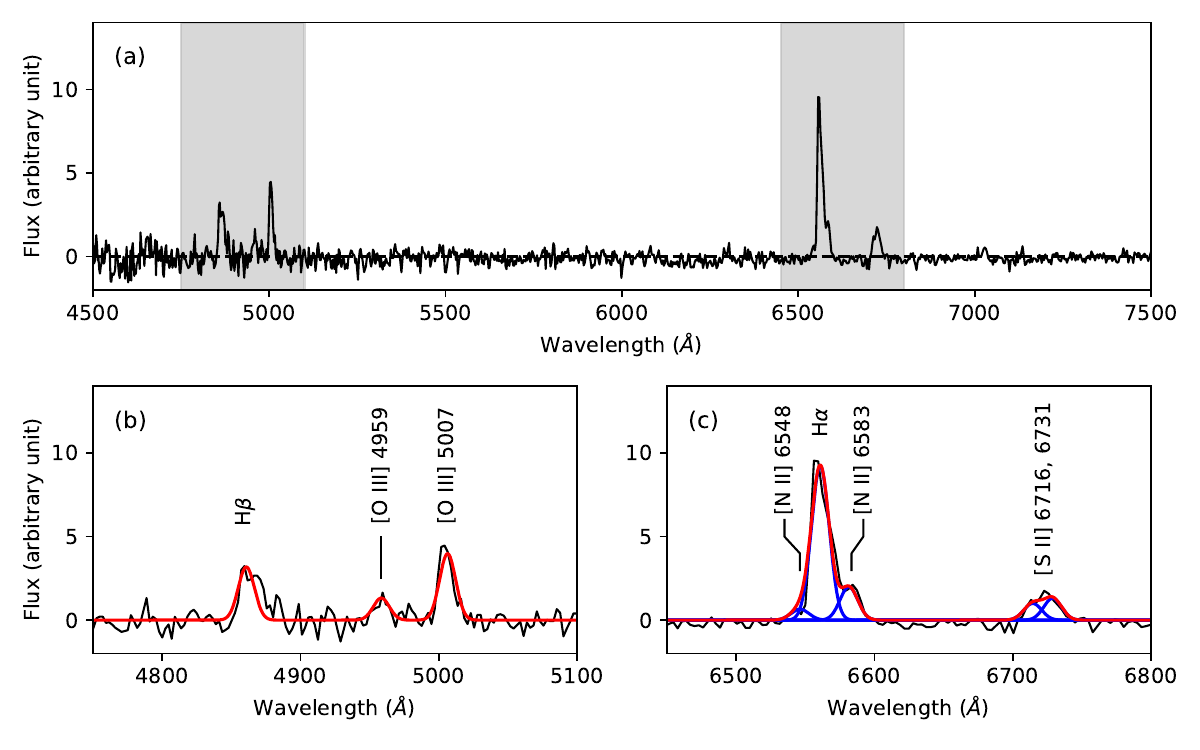}
    \caption{(a) Continuum-subtracted spectrum of the nearby H~\textsc{ii} region to the north of SN~2023ixf (Fig.~\ref{prephot}). The grey-shaded wavelength ranges are enlarged in panels (b) and (c), where the red and blue lines display the total and single Gaussian fits to the nebular emission lines, respectively.
    \label{spec.fig}} 
\end{figure*}

\begin{figure*}[htbp] 
    \centering 
    \includegraphics[width=0.9\linewidth]{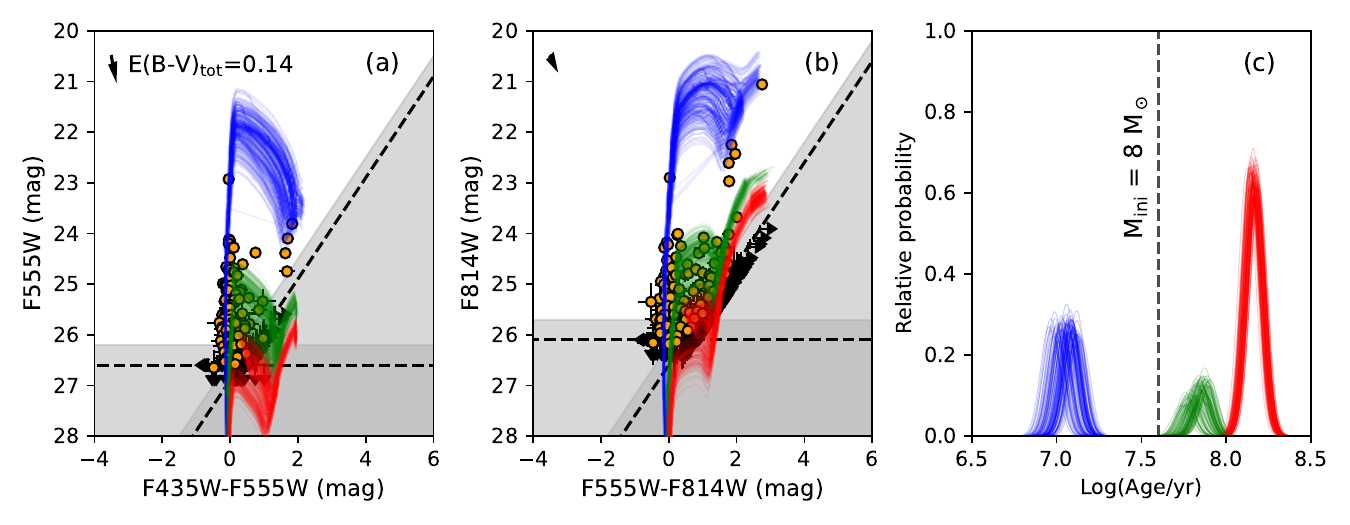}
    \caption{(a, b) Color-magnitude diagrams of resolved stellar populations within 150~pc from SN~2023ixf (orange data points) The error bars reflect their 1$\sigma$ photometric uncertainties. The dashed lines show the 50 per cent detection limits, and the grey-shaded regions show where $\leq$68 per cent of the artificial stars can be successfully recovered. Model stellar populations with three age components are fitted to the data, and the (blue, green, and red) thin lines are stellar isochrones from 100 random realizations according to the stellar log-age and extinction distributions of the model populations. The arrows in the upper-left corners are total (Galactic + internal) reddening vectors for a standard extinction law with $R_V$ = 3.1. (c) The star formation history in the vicinity of SN~2023ixf from 100 random realizations; the heights of the peaks are scaled to correspond to the weighting of each star formation epoch.
    \label{cmd.fig}} 
\end{figure*}

\subsection{Metallicity}\label{sec.chemical}

As shown in Fig.~\ref{prephot}, there are a few H~\textsc{ii} regions at distances of 150--300~pc to the north and west of SN~2023ixf. 
Part of the northern H~\textsc{ii} region is also covered by our long-slit follow-up spectroscopy of SN~2023ixf with the Xinglong 2.16-m telescope (Fig.~\ref{prephot}; Zhang et al. in preparation).  Here we use one of the acquired spectra to investigate the ionized gas in the SN environment. This spectrum was observed on June 1st by the Beijing Faint Object Spectrograph and Camera (BFOSC) and G4+385LP grism \citep{2016Fan}. The wavelength range is from 3700 $\AA$ to 8800 $\AA$ and the exposure time is 1200~s. The spectrum of the H~\textsc{ii} region was extracted from the red-outlined area in Fig.~\ref{prephot}, and data reduction was performed with the \textsc{astro-plpy}\footnote{https://pypi.org/project/astro-plpy} and \textsc{astro-wcpy} packages \citep{ruining_zhao_2023}.

Figure~\ref{spec.fig} shows the extracted spectrum, from which a stellar continuum has been fit (with the \textsc{ppxf} package; \citealt{Cappellari2004, Cappellari2017}) and removed. Prominent nebular emission lines are apparent (such as H$\alpha$, H$\beta$, [N~\textsc{ii}] $\lambda\lambda$ 6548, 6583, [O~\textsc{i}] $\lambda\lambda$ 6300, 6363, [O~\textsc{iii}] $\lambda\lambda$ 4959, 5007, and [S~\textsc{ii}] $\lambda\lambda$ 6716, 6731), and we measured their fluxes by fitting Gaussian profiles (Fig.~\ref{spec.fig}b, c). With the O3N2 calibration of the strong-line diagnositics \citep{Marino2013}, we derived an oxygen abundance of 12 + log(O/H) = 8.37 $\pm$~0.18~dex, which is lower than the solar value (8.69~dex; \citealt{Asplund2009}) by 0.32~dex (i.e. half-solar metallicity). We note, the strong-line method could suffer from possible systematic uncertainties. For example, the oxygen abundance would be 8.49 dex if adopting the empirical calibration in \citet{2004Pettini}. It is consistent with the above value within the margin of error.

\subsection{Star formation history}\label{sec.pop}

In the immediate SN vicinity, there are no obvious signs of ongoing star formation on the HST or IRAC images (e.g. young stellar complexes, H$\alpha$ emission or dust IR emission; Fig.~\ref{prephot}). In this area, the most recent star-forming activity may occur some time ago, after which star formation ceased or declined to very low levels. In order to recover the past star formation history, we analyzed the resolved stellar populations within $\sim$150~pc (typical scale of star-forming complexes; see \citealp{1995Efremov}) from SN~2023ixf based on their HST/ACS F435W/F555W/F814W photometry. In each band, we used only detections with signal-to-noise ratios larger than 5, and required their \textsc{dolphot} sharpness parameter to be in the range of $-$0.5 $<$ SHARP $<$ 0.5, so that the selected sources have point-like morphologies. We also used randomly positioned artificial stars to estimate a detection limit and additional photometric uncertainties induced by source crowding and imperfect sky subtraction. A total of 362 stars are detected in the local environment of SN~2023ixf and their color-magnitude diagrams are displayed in Fig.~\ref{cmd.fig}.

We then fitted model stellar populations (based on the \textsc{parsec} v1.2S stellar evolutionary models; \citealt{parsec.ref}) to the data with a hierarchical Bayesian method, which was detailed in \citet{Maund2016} and \citet{Sun2021} \citep[see also][]{Maund2017, Maund2018, Sun2020a, Sun2022a, Sun2023a, Sun2023b}. In brief, each model population has a \citet{imf.ref} initial mass function, a 50 per cent (non-interacting) binary fraction, and a flat distribution of secondary-to-primary mass ratio. Stars of each population has Gaussian log-age and extinction distributions, and we assumed a small log-age dispersion of 0.05~dex and a small extinction dispersion of 0.05~mag. Prolonged star formation can be considered as the superposition of multiple stellar populations with different mean log-ages.

We used three model populations to fit the data and solve for their parameters with the dynamic nested sampling package \textsc{dynesty} \citep{dynesty.ref}. We derived a mean (host galaxy) extinction of $A_V^{\rm host}$ = 0.4~mag, and mean log-ages of 7.07, 7.84 and 8.16~dex for the three age components (i.e. 12, 69, and 144~Myr). The derived star formation history and stellar isochrones corresponding to the three model populations are displayed in Fig.~\ref{cmd.fig}. Using more model populations may improve the accuracy of the derived star formation history but will not change the conclusion reached in this section. As later discussed, the SN progenitor corresponds to the youngest population and this population has already been fitted well.

We note that the derived extinction is significantly larger than that for the SN itself \citep[e.g.][]{2023Smith}. We argue that this is not unreasonable, since interstellar extinction often has significant spatial variation and the progenitor of SN~2023ixf could have expelled the nearby dust with its intensive radiation and stellar wind. Liu et al. (private communications) performed an environmental analysis of SN~2023ixf based on integral-field-unit spectroscopy. With Balmer decrement they found a total reddening of $E(B-V)_{\rm tot}$ = 0.11 $\pm$~0.06~mag for the ionized gas within 3~arcsec from SN~2023ixf. This corresponds to a host-galaxy extinction of $A_V^{\rm host}$ = 0.3 $\pm$~0.2~mag, consistent with our results.

Assuming single-star evolution, the most recent star-forming burst corresponds to a SN progenitor mass of $M_{\rm ini}$ = 17--19~$M_\odot$ (considering a conservative log-age uncertainty of 0.05~dex) and the earlier star formation epochs are too old to be consistent with a core-collapse SN. This result again suggests a relatively massive progenitor and is in agreement with the mass estimate derived from SED fitting (Section~\ref{sec.sedfitting}).

\section{Comparison with previous studies}\label{sec.comparison}

\citet{Pledgerixf} identified the progenitor of SN~2023ixf from the HST images and suggested it to be within the relatively low initial mass range of 8--10~$M_\odot$. They noted the star may be subject to significant extinction, which would be difficult to estimate without the near- and mid-IR data.

\citet{Kilpatrickixf} also reported the detection of the progenitor of SN~2023ixf. With SED fitting, they derived a progenitor mass of only $M_\odot$ = 11~$M_\odot$, which is significantly smaller than our result. This difference could be partly due to the stellar brightness variability. The HST observations were conducted at only one epoch, and it is difficult to estimate the phase-weighted average magnitudes in the optical bands. In order to account for this effect, we allowed the F675W and F814W model magnitudes to vary within 0.5~mag from the observed values, i.e. over a much larger range than the photometric uncertainties (Section~\ref{sec.sedfitting}). In fact, our best-fitting model SEDs predict brighter magnitudes in these two optical bands than the observed ones. In a test we found that the derived bolometric luminosity would be much smaller if we strictly required the model SEDs to match the observed F675W and F814W magnitudes within photometric uncertainties. 
This is consistent with the finding of \citet{Jencsonixf} that the HST observations were timed near the bottom of the pulsation cycle. In addition, we have included the $JHK$-band photometry reported by \citet{Soraisamixf}, while in \citet{Kilpatrickixf} the $JH$-band are detection limits. In our analysis we found the progenitor's SED peaks near the $J$ and $H$ bands (Fig.~\ref{hrd}), while their best-fitting model SED peaks at a slightly longer wavelength.

\citet{Jencsonixf} derived a bolometric luminosity of log($L/L_\odot$) = 5.1 $\pm$~0.2~dex and initial mass of $M_\odot$ = 17 $\pm$~4~$M_\odot$ for the progenitor of SN~2023ixf, which are roughly consistent with our results. 
Their analysis was performed by fitting the near-IR and mid-IR phase-weighted average magnitudes based on the \textsc{grams} models with O-rich silicate dust. 
As we pointed out in Section~\ref{sec.sedfitting}, however, the prominent silicate bump is rarely under the Spitzer/IRAC [8.0] detection limit. When a more strict limit (e.g 2$\sigma$) is applied, O-model is incompatible with the observations.
We prefer a C-rich dust chemical composition. 
We note that \citet{Jencsonixf} derived a detection limit of [8.0] $>$ 11.8~mag, significantly brighter than ours (14.95 mag). For comparison, the estimate of \citet{Kilpatrickixf} (their Table~2) corresponds to a more strict detection limit of [8.0] $>$ 16.1~mag, assuming a nominal zero-point flux of Vega of 64.9~Jy. The different values may arise from our different photometry techniques. 

\citet{Soraisamixf} accurately measured the progenitor's pulsational period and estimated its $K$-band absolute magnitude with the period-luminosity relation. They converted the absolute magnitude to a bolometric luminosity of log($L/L_\odot$) = 5.2--5.4~dex and inferred an initial mass of 20 $\pm$~4~$M_\odot$, both of which are slightly larger than ours. The period-luminosity relation they used was calibrated based on 255 RSGs in M31 \citep{PL.ref}. For the progenitor of SN~2023ixf, however, it is still unclear whether a RSG just before SN explosion would still follow the same period-luminosity relation as those at earlier evolutionary stages.

In summary, it is very challenging to derive accurate parameters for the progenitor of SN~2023ixf, due to its brightness variability, the uncertain circumstellar dust, and the poor understanding of the stellar evolutionary stage shortly before the explosion. The environmental analysis (Section~\ref{sec.env}) avoids these obstacles (although it has its own difficulties; see the discussion of \citealt{Sun2021}) and, as an independent analysis, has derived consistent results as our SED fitting (Section~\ref{sec.sedfitting}). We believe, therefore, that our conclusion should be reliable that SN~2023ixf has a relatively massive progenitor with an initial mass of $M_{\rm ini}$ = 16.2--17.4~$M_\odot$ (from SED fitting) or 17--19~$M_\odot$ (from SN environment).

\section{Summary and conclusions}

In this paper, we report a detailed analysis of the progenitor of the nearby Type~II SN~2023ixf. Two independent analyses, based on direct progenitor detection in pre-explosion observations and an analysis of the SN environment, are used and they reach consistent results.

The progenitor of SN~2023ixf is significantly detected on the pre-explosion images acquired by HST in the F658N, F675W, and F814W bands and by Spitzer in the [3.6] and [4.5] bands. In agreement with previous studies, the mid-IR light curves exhibit significant variability without obvious color changes.

The progenitor's SED is consistent with a RSG enshrouded by a dusty envelope. We modelled the SED by calculating the radiative transfer through dust; two different dust compositions were considered, i.e. C-rich pure graphite and O-rich pure silicate. 
Only the C-rich model seems consistent with observations and, assuming an effective temperature of $T_{\rm eff}$ $\sim$ 3700~K, the progenitor star has a bolometric luminosity of log($L/L_\odot$) = 5.11~dex, corresponding to an initial mass of $M_{\rm ini}$ = 16.2--17.4~$M_\odot$. The mass-loss rate is about 1 $\times 10^{-5} \ M_{\odot}$yr$^{-1}$. 

We also analyzed the environment of SN~2023ixf as another approach to understand its progenitor. A few H~\textsc{ii} regions are located at distances of 150--300~pc from the SN, and we derived a half-solar metallicity from strong nebular emission lines.

In the immediate SN vicinity ($<$150~pc) there are no obvious signs of ongoing star formation. We derived the star formation history based on the resolved stellar populations.  While most star-forming bursts are too old to be consistent with a core-collapse SN, the most recent one occurred 12~Myr ago, corresponding to an initial mass of $M_{\rm ini}$ = 17--19~$M_\odot$ for the progenitor of SN~2023ixf, assuming single-star evolution.

In summary, the progenitor of SN~2023ixf is among the most massive ones that have been directly probed for Type~II SNe. For such a massive progenitor, the powerful stellar wind likely drives significant mass loss and results in a low-mass H envelope, which could explain the relatively steep slope of the light curve out to 50~days after explosion \citep{Bianciardi2023}. It remains to be explored whether binary evolution plays any role for the progenitor of SN~2023ixf, although currently no obvious signs for a companion star have been discovered.

\section{acknowledgments}
We acknowledge the referee for his/her valuable comments and suggestions that improved the quality of the paper significantly. 
ZXN acknowledges the helpful discussions with Dr. Shu Wang, and NCS is grateful to Dr. Chenxu Liu for providing the extinction value estimated with itegral-field-unit spectroscopic data.
NCS's research is funded by the NSFC grant No.~12261141690.
The research of JRM is supported by the STFC consolidated grant ST/V000853/1.
JFL acknowledges support from the NSFC through grant Nos.~11988101 and 11933004, and support from the New Cornerstone Science Foundation through the New Cornerstone Investigator Program and the XPLORER PRIZE.

This research has made use of the NASA/IPAC Infrared Science Archive, which is funded by the National Aeronautics and Space Administration and operated by the California Institute of Technology.
This research is based on observations made with the NASA/ESA Hubble Space Telescope obtained from the Space Telescope Science Institute, which is operated by the Association of Universities for Research in Astronomy, Inc., under NASA contract NAS 5–26555. These observations are associated with program(s) 9490, 9720, and 6829.

Data used in this work are all publicly available from the NASA/IPAC Infrared Science Archive (https://sha.ipac.caltech.edu/applications/Spitzer/SHA/) and the Mikulski Archive for Space Telescope at the Space Telescope Science Institute via \dataset[10.17909/bhr8-jp04].


\bibliography{sample631}{}
\bibliographystyle{aasjournal}
\end{document}